\documentclass{pnastwo}
 
\usepackage[pdftex]{graphicx}
\usepackage{amssymb,amsfonts,amsmath}
\usepackage{pnastwoF}
\usepackage{bm}

\copyrightyear{2014}
\issuedate{Issue Date}
\volume{Volume}
\issuenumber{Issue Number}

\begin{document}

\title{Field-induced superconducting phase of FeSe in the BCS-BEC cross-over}

\author{
S.~Kasahara\affil{1}{Department of Physics, Kyoto University, Kyoto 606-8502, Japan}\affil{8}{These authors contributed equally to this work},
T.~Watashige\affil{1}{}\affil{2}{RIKEN Center for Emergent Matter Science, Wako, Saitama 351-0198, Japan}\affil{8}{}, 
T.~Hanaguri\affil{2}{},
Y.~Kohsaka\affil{2}{},
T.~Yamashita\affil{1}{},
Y.~Shimoyama\affil{1}{},
Y.~Mizukami\affil{1}{}\affil{3}{Department of Advanced Materials Science, University of Tokyo, Chiba 277-8561, Japan},
R.~Endo\affil{1}{},
H.~Ikeda\affil{1}{}\affil{7}{Present address: Department of Physical Sciences, Ritsumeikan University, Kusatsu, Shiga 525-8577, Japan},
K.~Aoyama\affil{1}{}\affil{4}{The Hakubi Center for Advanced Research, Kyoto University, Kyoto 606-8501, Japan}, 
T.~Terashima\affil{5}{National Institute for Materials Science, 3-13 Sakura, Tsukuba, Ibaraki, 305-0003 Japan}, 
S.~Uji\affil{5}{},
T.~Wolf\affil{6}{Institute of Solid State Physics (IFP), Karlsruhe Institute of Technology, D-76021 Karlsruhe, Germany}, 
H.~v. L\"ohneysen\affil{6}{},
T.~Shibauchi\affil{1}{}\affil{3}{},
and
Y.~Matsuda\affil{1}{}\affil{9}{To whom correspondence should be addressed. Email: {\sf \small matsuda@scphys.kyoto-u.ac.jp}}}


\maketitle

\begin{article}
\begin{abstract}
{Fermi systems in the crossover regime between weakly coupled Bardeen-Cooper-Schrieffer (BCS) and strongly coupled Bose-Einstein-condensate (BEC) limits are among the most fascinating objects to study the behavior of an assembly of strongly interacting particles. The physics of this crossover has been of considerable interest  both in the fields of condensed matter and ultracold atoms. One of the most challenging issue in this regime is the effect of large spin imbalance on a Fermi system under magnetic fields. Although several exotic physical properties have been predicted theoretically, the experimental realization of such an unusual superconducting state has not been achieved so far. Here we show that pure single crystals of superconducting FeSe offer the possibility to enter the previously unexplored realm where the three energies, Fermi energy $\varepsilon_{\rm F}$, superconducting gap $\Delta$ and Zeeman energy, become comparable.  Through the superfluid response, transport, thermoelectric response, and spectroscopic-imaging scanning tunneling microscopy, we demonstrate that $\varepsilon_{\rm F}$ of FeSe is extremely small, with the ratio $\Delta/\varepsilon_{\rm F}\sim1~(\sim0.3)$ in the electron (hole) band.   Moreover, thermal-conductivity measurements give evidence of a distinct phase line below the upper critical field, where the Zeeman energy becomes comparable to $\varepsilon_{\rm F}$ and $\Delta$.  The observation of this field-induced phase provides insights into previously poorly understood aspects of the highly spin-polarized Fermi liquid in the BCS-BEC crossover regime.}
\end{abstract}


\abbreviations{BCS, Bardeen-Cooper-Schrieffer; BEC, Bose-Einstein condensation; ARPES, angle-resolved photoemission spectroscopy; QPI, quasiparticle interference}
 
{\bf 
\noindent
Significance\\
The BCS-BEC crossover bridges the two important theories of bound particles in a unified picture with the ratio of the attractive interaction to  the Fermi energy as a tuning parameter. A key issue is to understand the intermediate regime, where new states of matter may emerge. Here, we show that the Fermi energy of FeSe is extremely small, resulting in that this system can be regarded as an extraordinary ``high-temperature" superconductor located at the verge of a BCS-BEC crossover. Most importantly, we discover the emergence of an unexpected superconducting phase in strong magnetic fields, demonstrating that the Zeeman splitting comparable to the Fermi energy leads to a strong modification of the properties of fermionic systems in such a regime.
\\
}

\dropcap{S}uperconductivity in most metals is well explained by the weak-coupling BCS theory, where the pairing instability arises from weak attractive interactions in a degenerate fermionic system. In the opposite limit of BEC, composite bosons consisting of strongly coupled fermions condense into a coherent quantum state \cite{Nozi85,Legg80}.  In BCS superconductors, the superconducting transition temperature is usually several orders of magnitude smaller than the Fermi temperature, $T_{\rm c}/T_{\rm F}=10^{-5}$-$10^{-4}$, while in the BEC limit $T_{\rm c}/T_{\rm F}$ is of the order of $10^{-1}$.  Even in the high-$T_{\rm c}$ cuprates, $T_{\rm c}/T_{\rm F}$ is merely of the order of $10^{-2}$ at optimal doping. Of particular interest is the BCS-BEC crossover regime with intermediate coupling strength. In this regime the size of interacting pairs ($\sim \xi$), which is known as the coherence length, becomes comparable to the average distance between particles ($\sim 1/k_{\rm F}$), i.e., $k_{\rm F}\xi\sim 1$ \cite{Chen05,Gior08,Rand14}, where $k_{\rm F}$ is the Fermi momentum. This regime is expected to have the highest values of $T_{\rm c}/T_{\rm F}=0.1-0.2$ and $\Delta/\varepsilon_{\rm F} \sim 0.5$ ever observed in any fermionic superfluid. 

One intriguing issue concerns the role of spin imbalance, whether it will lead to a strong modification of the properties of the Fermi system in the crossover regime.  This problem has been of considerable interest not only in the context of superconductivity but also in ultracold-atom physics \cite{Che10,Shin07,Gub12}.    However,  such Fermi systems have been extremely hard to access.  In superconductors, the spin imbalance can be introduced through the Zeeman effect by applying a strong magnetic field.   Again,  in the high-$T_{\rm c}$ cuprates, the Zeeman energy at the upper critical field at $T\ll T_{\rm c}$ is of the order of only  $10^{-2}\varepsilon_{\rm F}$.  In ultracold atoms, although several exotic superfluid states have been proposed \cite{Guba03,Yosh07}, cooling the systems down to sufficiently low temperature ($T\ll T_{\rm c}$) is not easily attained.

FeSe provides an ideal platform for studying a highly spin-polarized Fermi system in the crossover regime.  FeSe is the simplest iron-based layered superconductor (Inset of Fig.\;1{\it A}) with $T_{\rm c}$ of $\sim$9\,K \cite{Hsu08}. 
The structural transition from tetragonal to orthorhombic crystal symmetry occurs at $T_{\rm s}\approx$ 90\;K and a large electronic in-plane anisotropy appears.  In contrast to the other iron-based compounds, no magnetic order occurs below $T_{\rm s}$.   A prominent feature of the pseudobinary ``11" family (FeSe$_{1-x}$Te$_x$) is the presence of very shallow pockets, as reported by angle-resolved photoemission-spectroscopy (ARPES). Although a possible BCS-BEC crossover has been suggested in the bands around the $\Gamma$-point \cite{Luba12,Okaz13}, it is still an open question whether all bands are in such a crossover regime.  Moreover, it should be noted that high-quality single crystals are highly requisite for the study of the crossover regime, as exotic superconductivity often is extremely sensitive to impurities. Previous FeSe$_x$ single crystals are strongly disordered, as indicated by large residual resistivity $\rho_0$ and small residual resistivity ratio $RRR$, typically 0.1 m$\Omega$cm and $\sim$\;5, respectively \cite{Brai09}. 

\section{Results and Discussion}
\subsection{BCS-BEC crossover}

By using high-quality single crystals of FeSe ({\it SI Text 1}, Figs.\;S1 and S2) which have become available recently, we have measured the transport properties (Fig.\;1{\it A}).    In zero field, the temperature dependence of the resistivity $\rho$ can be described by $\rho=\rho_0+AT^{\alpha}$ with $\alpha=1.05-1.2$ below 25\,K. Taking $\rho(T_{\rm c}^+) \approx 10\; \mu \Omega$cm as upper limit of $\rho_0$ leads to $RRR > 40$, i.e., a factor of 10 higher than previous samples.   In the present crystals $T_c$ defined by the zero resistivity is 9.5\;K, which is higher than $T_c\sim 8$\;K of the low $RRR$ samples \cite{Brai09}.     A remarkably large magnetoresistance (Inset of Fig.\;1{\it A})  not observed in previously studied low-$RRR$ crystals \cite{Brai09}, supports that the crystal is very clean ({\it SI Text 2}, Fig.\;S3{\it A}). The strongly $T$-dependent Hall coefficient $R_{\rm H}$ below $\sim$\;60\;K indicates that the electron and hole mobilities are of the same order  ({\it SI Text 2}, Fig.\;S3{\it B}).  The London penetration depth $\lambda_{\rm L}$ shows a quasi $T$-linear dependence, $\lambda_{\rm L}(T)\propto T^{1.4}$,  for $T/T_{\rm c}<0.2$,  suggesting the presence of line nodes in the superconducting gap (Fig.\;1{\it B}). Figure\;1{\it C} shows the tunneling conductance which is proportional to the density of states, measured with a scanning tunneling microscope at 0.4\;K. The V-shaped spectrum at low bias voltages likewise indicates the presence of line nodes, which is consistent with previously reported observations \cite{Song11}.  We note that the line nodes are accidental, not symmetry protected, i.e., the gap function is extended $s$-wave, because the nodes are absent in samples with low $RRR$ \cite{Dong09}.  Distinct peaks and shoulder structures in the spectra indicate the presence of (at least) two superconducting gaps ($\Delta \approx$ 2.5 and 3.5\;meV), reflecting the multiband nature.  

\begin{figure}[t]
\centerline{\includegraphics[width=.9\linewidth]{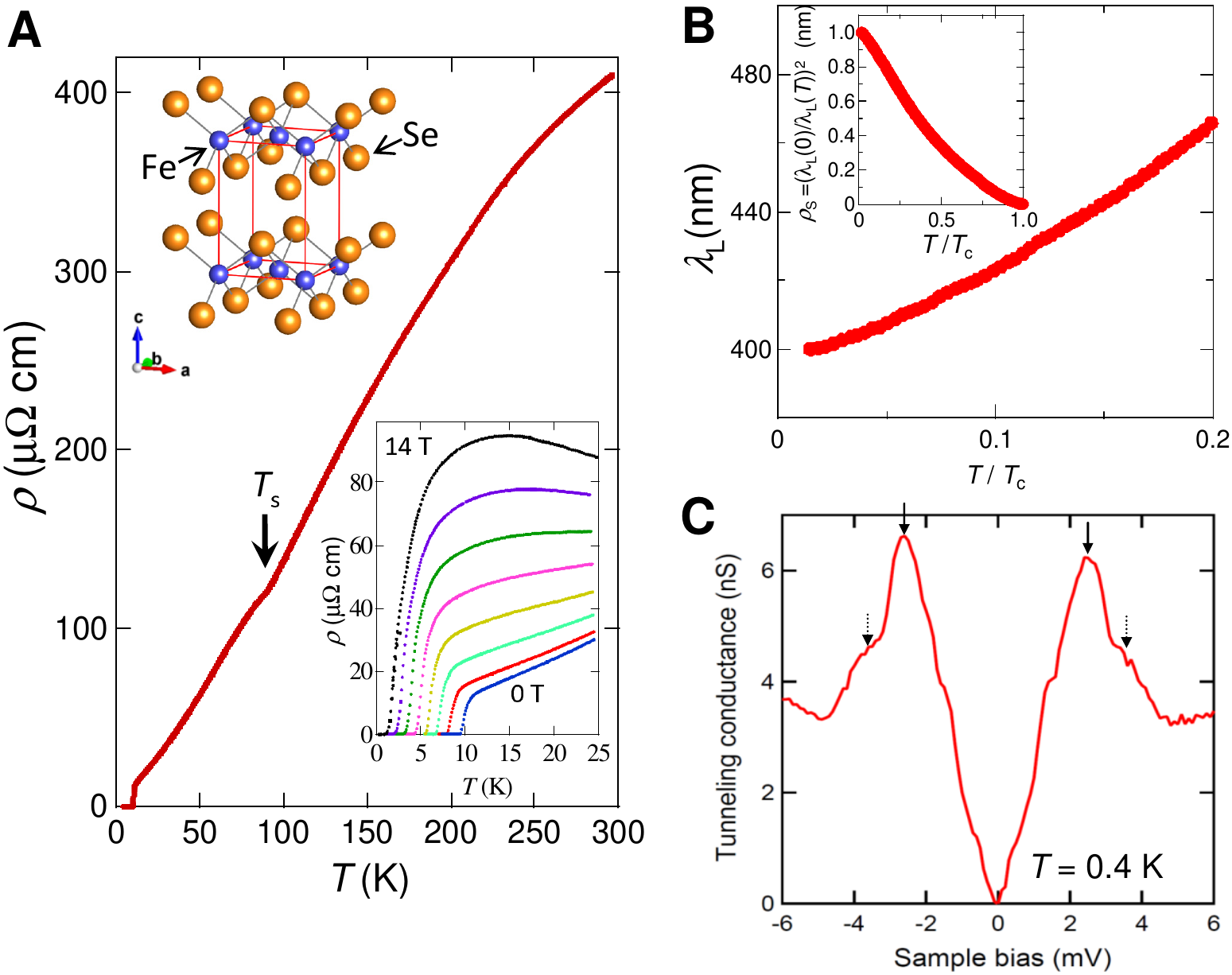}}
\caption{
Normal and superconducting properties of high quality single crystals of FeSe. ({\it A}): Temperature dependence of the in-plane resistivity$\rho$ of FeSe. The structural transition occurs at $T_{\rm s}\approx$\;90\;K. The upper inset shows the crystal structure.  The lower inset shows $\rho(T)$ in magnetic field ($\bm{H}\parallel c$). From bottom to top,  $\mu_0H =$ 0, 2, 4, 6, 8, 10, 12, and 14\;T is applied. ({\it B}): Temperature dependence of the London penetration depth. Inset shows the superfluid density normalized by the zero temperature value  $\rho_{\rm s}\equiv\lambda_{\rm L}^2(0)/\lambda_{\rm L}^2(T)$. ({\it C}): Tunneling conductance spectrum at $T =0.4$\;K. The peaks at $\pm2.5$\;meV (arrows) and shoulder structures at $\pm3.5$\;meV (dashed arrows) indicate the multiple superconducting gaps. 
}\label{fig1}
\end{figure}

The high quality of the single crystals enables us to estimate the Fermi energies  $\varepsilon_{\rm F}^{\rm e}$ and $\varepsilon_{\rm F}^{\rm h}$ from the band edges of electron and hole sheets, respectively, by using several techniques; all of them consistently point to extremely small Fermi energies.    In 2D systems $\varepsilon_{\rm F}$ is related to $\lambda_{\rm L}(0)$ as $\varepsilon_{\rm F}=\frac{\pi \hbar^2 d}{\mu_0e^2}\lambda_{\rm L}^{-2}(0)$, where $d$ is the interlayer distance and $\mu_0$ is the vacuum permeability. From the $T$ dependence of $\lambda_{\rm L}(T)$, we obtain $\lambda_{\rm L}(0) \approx 400$~nm (Fig.\;1{\it B}, {\it SI Text 3}, Fig.\;S4) \cite{Hash12}. 
Very recent quantum oscillation measurements on the present FeSe crystals revealed that the Fermi surface consists of one hole sheet and one (compensating) electron sheet (see Fig.\;2{\it A}) \cite{Tera14}.  Then $\lambda_{\rm L}$ can be written as $1/\lambda_{\rm L}^2=1/(\lambda_{\rm L}^{\rm e})^2+1/(\lambda_{\rm L}^{\rm h})^2$, where $\lambda_{\rm L}^{\rm e}$ and $\lambda_{\rm L}^{\rm h}$ represent the contribution from the electron and hole sheets, respectively.   Assuming that two sheets have similar effective masses as indicated by the Hall effect (see below and {\it SI Text 2}), we estimate $\varepsilon_{\rm F}^{\rm e}\sim \varepsilon_{\rm F}^{\rm h} \sim 7-8$\;meV.   The magnitude of the Fermi energy can also be inferred from the thermoelectric response in the normal state  ({\it SI Text 2}).  From the Seebeck coefficient $S$, the upper limit of $\varepsilon_{\rm F}^{\rm e}$ is deduced to be $\sim$ 10\;meV ({\it SI Text 2}, Fig.\;S3{\it C}).  Moreover, the sign change of  $R_{\rm H}(T)$  at 60\;K  ({\it SI Text 2}, Fig.\;S3{\it B}) indicates that the Fermi energies $\varepsilon_{\rm F}^{\rm e}$ and $\varepsilon_{\rm F}^{\rm h}$ are of similar size, a feature also observed in underdoped cuprate superconductors with small electron and hole pockets \cite{LeBo11}. In contrast to the cuprate case, however, $R_{\rm H}$ in FeSe almost vanishes at high temperatures, which sheds light on the unique feature of FeSe with extremely small Fermi energy.

We can determine the electron and hole Fermi energies directly by measuring the electronic dispersion curves in momentum space yielding $\varepsilon_{\rm F}^{\rm h} = E_{\rm HT}-E_{\rm F}$ ($\varepsilon_{\rm F}^{\rm e} = E_{\rm F}-E_{\rm EB}$) for the hole (electron) band. Here $E_{\rm HT}$ ($E_{\rm EB}$) is the energy of the top (bottom) of the hole (electron) band and $E_{\rm F}$ is the electrochemical potential.   For this assignment, we exploit spectroscopic-imaging scanning tunneling microscopy to observe the quasiparticle interference (QPI) patterns~\cite{Hase93,Crom93} associated with electron waves scattered off by defects. By taking the Fourier transform of energy-dependent normalized conductance images, characteristic wave vectors of electrons at different energies reflecting the band dispersion, can be determined ({\it SI Text 4}). The observed QPI patterns of FeSe at 1.5\;K ({\it SI Text 4}, Fig.\;S5) consist of hole- and electron-like branches that disperse along the crystallographic $b$ and $a(<b)$ directions, respectively. These branches can naturally be ascribed to the hole and electron sheets illustrated  in Fig.\;2{\it A}.  The QPI signals exhibit a strong in-plane anisotropy.  Such an anisotropy is consistent with the largely elongated vortex core structure~\cite{Song11}.   The origin of the strong in-plane anisotropy of the QPI signals is unclear,  but a possible cause may be the orbital ordering in the orthorhombic phase.   

\begin{figure*}[b]
\centerline{\includegraphics[width=0.705\textwidth]{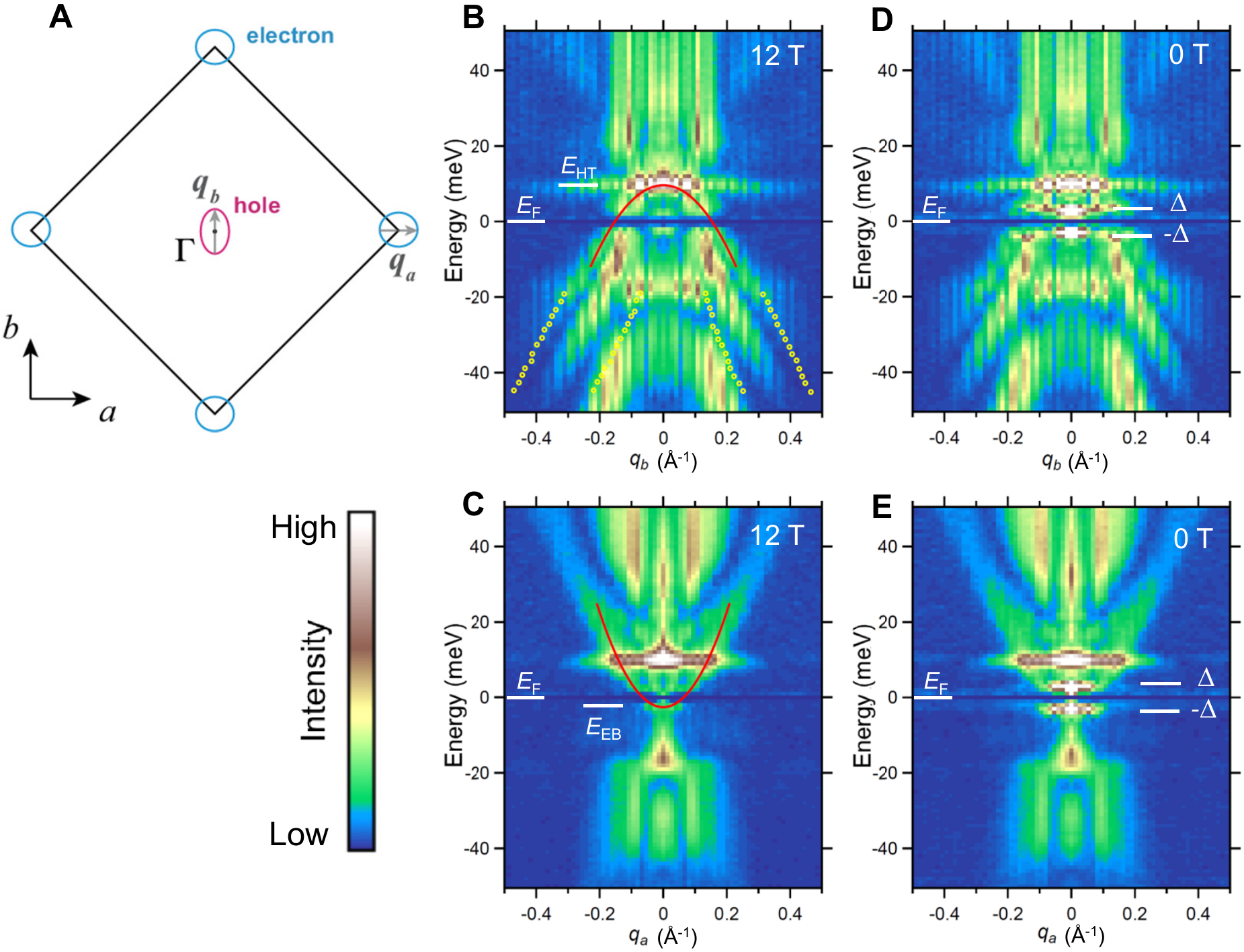}}
\caption{
The band dispersions obtained from the QPI pattern. ({\it A}): Schematic figure of the Fermi surface of FeSe inferred from QPI.    ({\it B,\,C}): QPI dispersions at 12\;T obtained by taking linecuts from the energy-dependent Fourier-transformed normalized conductance images ({\it SI Text 4}, Fig.\;S5) along $q_a$ and $q_b$, respectively. A $q$-independent feature at $\sim$+10\;meV is associated with defect states ({\it SI Text 4}, Fig.\;S6). Peak positions of the representative branches are fitted with parabolic function to obtain Fermi energies and effective masses (solid lines).
The top (bottom) of the hole (electron) band $E_{\rm HT}$ ($E_{\rm EB}$) are indicated by white bars. 
Expected intra-band scattering vectors associated with the $\alpha$-band detected by ARPES \cite{Male13} are plotted in ({\it B}) by yellow circles. ({\it D,\,E}):  QPI dispersions at $H = 0$. A pair of sharp intensity peaks ($\pm \Delta$) appears at $E \approx \pm2$\;meV due to the opening of the superconducting gap. Superconducting gap defined by the positions of the coherence peaks (white bars) are comparable to the Fermi energies. Note that the feature attributed to defect states is independent of field.
}\label{fig2}
\end{figure*}

As shown in Figs.\;2{\it B} and 2{\it C}, full dispersion curves of hole- and electron-like branches are clearly identified by taking linecuts from the series of Fourier-transformed conductance images ({\it SI Text 4}, Fig.\;S5). Here, a magnetic field $\mu_0 H = 12$\;T is applied parallel to the $c$ axis ($\bm{H}\parallel c$) to mostly suppress superconductivity.   Multiple hole-like branches are identified in Fig.\;2{\it B}.    Since QPI signals include both intra- and interband scattering processes, it is difficult to disentangle all the QPI branches in order to resolve the bare band structure.   Nevertheless, the top of the hole band can be faithfully determined to yield $\varepsilon_{\rm F}^{\rm h} \sim$ 10\,meV from the highest energy of the top-most branch. This branch is quantitatively consistent with the intra-band scattering associated with the $\alpha$-band detected by ARPES \cite{Male13}.   In the case of the electron-like branch, an even smaller band bottom of $\varepsilon_{\rm F}^{\rm e} \sim$ 2 \,--\,3\,meV is estimated (Fig.\;2{\it C}).   These small values are consistent with the ones estimated from the superfluid and thermoelectric responses.  The effective mass of electron (hole) determined by QPI assuming parabolic dispersion is  2.5\;$m_0$ (3.5\;$m_0$), where $m_0$ is the free-electron mass.  The observation of comparable effective masses of electrons and holes is consistent with the Hall-effect data ({\it SI Text 2}).  We stress that the electronic structure obtained  from QPI, including masses of electron and hole,  the size and the number of each pocket, and the magnitude of the Fermi energies, is consistent with the values recently reported by the quantum oscillations in the quantitative level \cite{Tera14}.  Remarkably,  the superconducting gaps  are of the same order as the Fermi energy of each band, $\Delta/\varepsilon_F^e\sim 1$ (Fig.\;2{\it D}) and  $\Delta/\varepsilon_F^h\sim 0.3$ (Fig.\;{2\it E}), implying the BCS-BEC crossover regime. Additional strong support of the crossover is provided by extremely small $k_F\xi \sim$ 1\,--\,4. Here $k_F$ of the electron (hole) sheet obtained from Fig.\;2{\it C} (Fig.\;2{\it B}) is roughly 0.3 (0.75)\;nm$^{-1}$, and $\xi$  determined from the upper critical field ($\sim 17$\;T) is roughly 5\;nm.

\subsection{Field-induced superconducting phase}

So far we discussed the relation between $\varepsilon_{\rm F}$ and $\Delta$.  How does the Zeeman  energy scale $\mu_{\rm B}H$, where $\mu_{\rm B}$ is the Bohr magneton, enter the game?  The thermal conductivity $\kappa$ is well suited to address the issue of how the magnetic field affects the extraordinary pairing state by probing quasiparticle excitations out of the superconducting condensate,  as the Cooper pair condensate does not contribute to heat transport.    Figure\;3{\it A} shows the $T$ dependence of  $\kappa/T$ in zero field.   Below $T_{\rm c}$, $\kappa$ is enhanced due to the suppression of quasiparticle scattering rates owing to the gap formation.   As shown in the inset of Fig.\;3{\it A}, $\kappa/T$ at low temperatures is well fitted as $\kappa/T=\kappa_0/T+\beta T$, similar to Tl$_2$Ba$_2$CuO$_{6+\delta}$~\cite{Haw07}.  The presence of the residual $\kappa_0/T$ at $T\rightarrow 0$ is consistent with line nodes in the gap.

 Figure\;3{\it B} shows the $H$ dependence of $\kappa/T$ for $\bm{H}\parallel c$ well below $T_{\rm c}$ obtained after averaging over many field sweeps at constant temperatures.     Beyond the initial steep drop at low fields, likely caused by the suppression of the quasiparticle mean free path $\ell_{\rm e}$ through the introduction of vortices,  $\kappa(H)/T$ becomes nearly $H$-independent.  Similar behavior has been reported for Bi$_2$Sr$_2$CaCu$_2$O$_8$ \cite{Kris97} and CeCoIn$_5$ \cite{Kasa05}.  It has been suggested  that 
the nearly $H$-independent $\kappa$ reflects a compensation between the enhancement of the density of states by magnetic field in nodal superconductors (Doppler shift) and the concomitant reduction in $\ell_{\rm e}$ due to increased scattering from vortices \cite{Fran82}.  
At high fields, above the smoothly varying background, $\kappa(H)/T$ exhibits a cusp-like feature at a field $H^{\ast}$  that is practically independent of $T$.  The height of the cusp-like peak decreases with increasing $T$. To further analyze our data,  the $\kappa/T$ values at different temperatures are extrapolated  to $T=0$ for each field value measured to yield $\kappa_0(H)/T$  as shown in Fig.\;3{\it C}, corroborating the robustness of the cusp.  In particular, the cusp of $\kappa_0/T$ is unrelated to phonon heat transport because  phonons do not contribute to $\kappa/T$ for $T\rightarrow0$.  Since the thermal conductivity has no fluctuation corrections \cite{Vish01}, the cusp of $\kappa$ usually corresponds to the mean-field phase transition.  We note that at $H^*$ the field dependence of magnetic torque shows no discernible anomaly ({\it SI Text 5}, Fig.\;S7).  However, such a difference of the sensitivity to the transition in different measurements has been reported for the field-induced transition between two superconducting phases in CeCoIn$_5$, which is hardly resolved in magnetization\cite{Taya02}, despite clear anomaly in some other bulk probes\cite{Mats07}. Moreover, the hysteresis in the magnetization due to vortex pinning may smear out a possible  torque anomaly at $H^*$.

\begin{figure}[b!]
\centerline{\includegraphics[width=.95\linewidth]{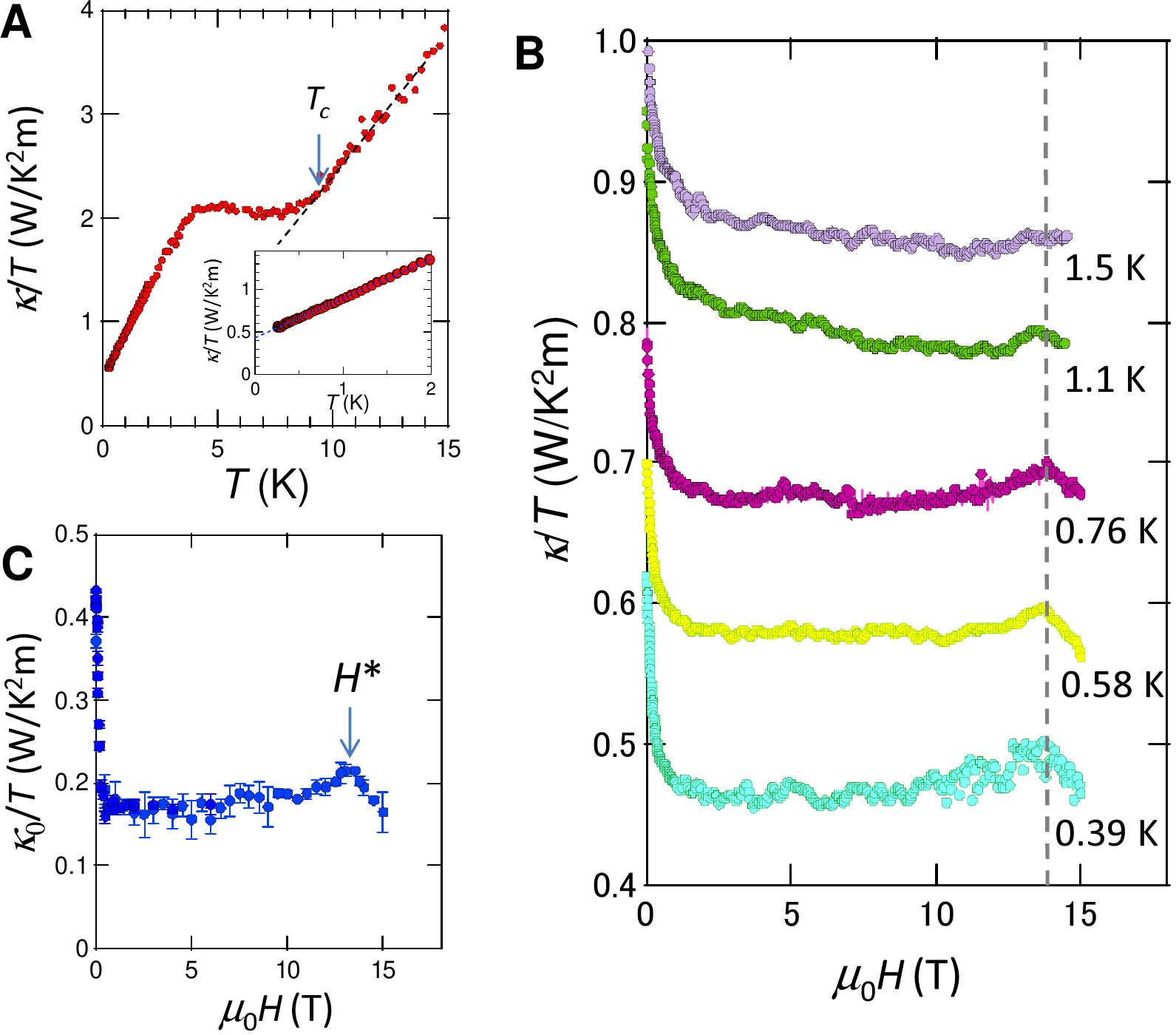}}
\caption{
Field induced transition revealed by the thermal conductivity. 
({\it A}): Temperature dependence of the in-plane thermal conductivity divided by temperature, $\kappa/T$.  Arrow marked $T_{\rm c}$ indicates the onset temperature of the superconductivity determined by zero resistivity (Fig.\;1{\it A}) and zero thermoelectric power ({\it SI Text 2}, Fig.\;S3{\it C}).  The inset shows  $\kappa/T$ at low temperatures.  ({\it B}):  Magnetic-field dependence of $\kappa/T$ at low temperatures ($\bm{H}\parallel c$). No hysteresis with respect to the field-sweep direction is observed.  $\kappa/T$ shows a plateau-like behavior in a wide field range.  At $H^{\ast}$, $\kappa/T$ shows a cusp-like peak, suggestive of a nearly temperature-independent transition (dashed line).  At 1.5\;K, the cusp disappears and a weak structure (within the accuracy of the measurement) is observed at lower field.  ({\it C}): Magnetic-field dependence of $\kappa/T$ in the zero-temperature limit, $\kappa_0/T$, obtained by linear extrapolation of $\kappa/T$ versus $T$ at low temperatures
}\label{fig3}
\end{figure}

Figure\;4 displays the $H$-$T$ phase diagram for $\bm{H}\parallel c$.  The irreversibility fields $H_{\rm irr}$ at low temperatures extend to fields well above $H^{\ast}$, indicating that $H^{\ast}$ is located inside the superconducting state.   The anomaly at $H^{\ast}$ is not caused by some changes of the flux-line lattice, such as melting transition, because the peak field determined by the torque is strongly $T$-dependent and well below $H^{\ast}$ ({\it SI Text 5}, Fig.\;S7), indicating that the flux-line lattice is already highly disordered at $H^{\ast}$. 

As shown in the inset of Fig.\;1{\it A}, the resistivity at $\mu_0H=14$\;T increases with decreasing temperature and decreases after showing a broad maximum at around 15\;K.  The $T$-dependence at high temperature  is a typical behavior of the very pure compensated semimetals.   However, the decrease of the resistivity at  low temperature regime is not expected in conventional semimetals.  This unusual decrease may be attributed to the  strong superconducting fluctuations above $H_{\rm irr}$ (inset of Fig.\;4).    In higher fields  the fluctuation region expands to higher $T>T_{\rm c}$.   In fact,   the Ginzburg number, which is given by  $G_{\rm i}\sim (T_{\rm c}/T_{\rm F})^4$ within the BCS framework \cite{Larkin},  is orders of magnitude larger than in any other superconductors.    This large range of fluctuations may be related to the presence of preformed pairs predicted in the BCS-BEC crossover regime \cite{Nozi85,Chen05,Gior08,Rand14}.

The appearance of the high-field phase (B-phase in Fig.\;4) where three characteristic energy scales are comparable, $\mu_{\rm B}H^{\ast}\sim \varepsilon_{\rm F}\sim \Delta(0)$,  suggests a phase transition of the Fermi liquid  with strong spin imbalance  in the BCS-BEC crossover regime.   Whether the observed distinct phase arises from strong spin imbalance and/or a BCS-BEC crossover, however, needs to be resolved in the future with particular attention to multiband effects.  We discuss two possible scenarios.  (1) The  phase line might signal an electronic transition akin to a Lifshitz transition, i.e., a topology change of the Fermi surface.  Indeed the phase line would be independent of $T$ and smeared by thermal fluctuations.  However, the fact that this phase line vanishes at $H_{\rm irr}$ would be accidental.  Furthermore, the absence of any discernible anomaly in torque magnetometry at $H^{\ast}$ ({\it SI Text 5}, Fig.\;S7) implies that the $\kappa(H)/T$ anomaly at $H^{\ast}$  is not caused by a Lifshitz transition nor, for that matter, by a spin-density-wave type of magnetic order.  (2) Comparable Fermi and Zeeman energies may  lead to an unprecedented superconducting state of highly spin-polarized electrons, such as  spin-triplet pairing and an admixture of even- and odd-frequency pairing \cite{Yana08}.  
Comparable gap and Zeeman energies may alternatively induce  a Fulde-Ferrell-Larkin-Ovchinnikov (FFLO)-like state with Cooper pairs having finite total momentum ($\bm{k}\uparrow$, $-\bm{k}+\bm{q}\downarrow$)  owing to the pairing channel between the Zeeman-split Fermi surfaces \cite{Mats07}.  The FFLO state requires a large Maki parameter, i. e., a ratio of orbital and Pauli-paramagnetic limiting fields, $\alpha_{\rm M}\equiv \sqrt{2} {H_{\rm c2}^{\rm orb}}/{H_{\rm c2}^{\rm P}}> 1.5$ in the BCS limit. In this regime, $\alpha_{\rm M} \approx 2 m^{\ast}/{m_0}\cdot {\Delta}/{\varepsilon_{\rm F}}$, yielding, for FeSe,  $\alpha_{\rm M}$ as large as $\sim 5$ ($\sim 2.5$) in the electron (hole) pockets.  This estimate may be questionable in the regime of ${\Delta}/\varepsilon_{\rm F}$ for FeSe. In any case, we stress that the high-field phase is not a simple FFLO phase because in the multiband superconductor FeSe the interaction between  electron and hole bands is crucial. 
Even in the single band systems, it has been suggested that the FFLO state  becomes unstable in the crossover regime \cite{Rad10}.   Our work should motivate further studies in the field of  strongly interacting Fermi liquids near the BCS-BEC crossover regime and in the presence of large spin imbalance, which remains largely unexplored and might bridge the areas of condensed-matter and ultracold-atom systems.
 
\begin{figure}[h]
\centerline{\includegraphics[width=.75\linewidth]{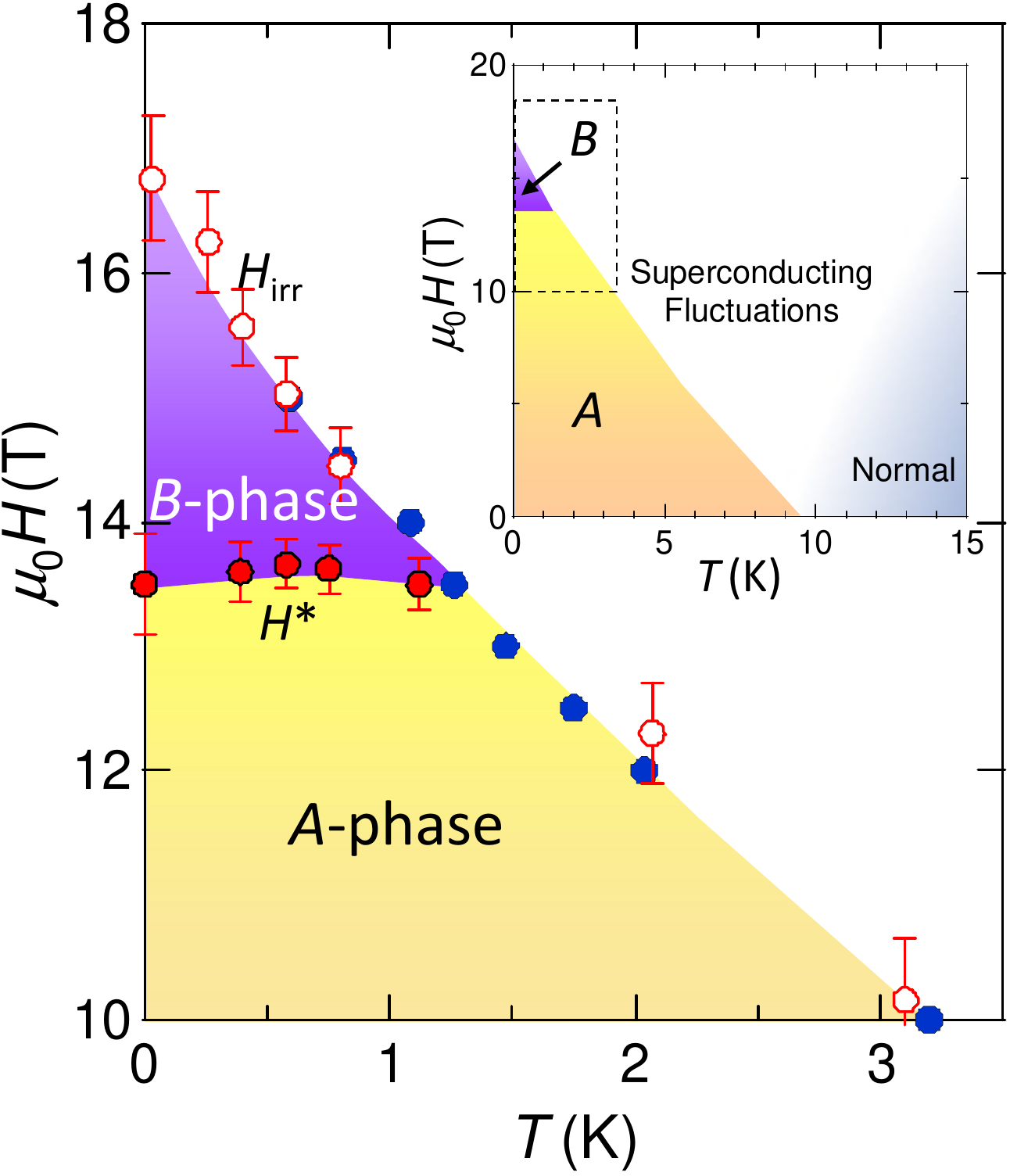}}
\caption{
$H$-$T$ phase diagram of FeSe for $\bm{H} \parallel c$.  
Solid blue and open red circles represent $H_{\rm irr}$ determined by zero resistivity and the onset of  hysteresis loops of the magnetic torque, respectively.  The mean-field upper critical field is above $H_{\rm irr}$.   Solid red circles represent $H^{\ast}$ determined by the cusp of $\kappa(H)/T$ shown in Figs.\;3({\it B}) and ({\it C}), separating a high-field superconducting phase ($B$-phase) from the low-field phase ($A$-phase).   The inset shows the overall $H$-$T$ phase diagram.   Superconducting fluctuation regime is determined by the temperature at which the resistivity deviates from the behavior expected in conventional semimetallic systems.
}\label{fig4}
\end{figure}

\begin{acknowledgments}
We thank A.\,E.~B\"ohmer,  A. V. Chubukov, I. Eremin, P. J. Hirschfeld, H. Kontani, S. S. Lee, C.~Meingast, A. Nevidomskyy, L. Radzihovsky, M. Randeria, I. Vekhter, and Y. Yanase for valuable discussion. This work has been supported by Japan -- Germany Research Cooperative Program, Grant-in-Aid for Scientific Research (KAKENHI) from Japan Society for the Promotion of Scienceand Project No. 56393598 from German Academic Exchange Service, and the ``Topological Quantum Phenomena" (No. 25103713) KAKENHI on Innovative Areas from Ministry of Education, Culture, Sports, Science and Technology of Japan.
\end{acknowledgments}

\end{article}



\clearpage

\renewcommand{\figurename}{Fig. S$\!\!$}
\renewcommand{\theequation}{S\arabic{equation}}
\renewcommand{\thetable}{S\arabic{table}}
\setcounter{figure}{0}



\begin{article}

\noindent
{\huge\bf Supporting Information} 

\bigskip
\noindent
{\large\bf Kasahara {\it et al.} }


\section{SI1--Sample growth and characterization}

High-quality single crystals of tetragonal $\beta$-FeSe were grown by the vapor transport method at Karlsruhe Institute of Technology~\cite{Boehmer13}.  A mixture of Fe and Se powders was sealed in an evacuated SiO$_2$ ampoule together with KCl and AlCl$_3$ powders. The ampoule was heated to 390$^\circ$C on one end while the other end was kept at 240$^\circ$C. After 28.5 days,  single crystals of FeSe with tetragonal morphology were extracted at the cold end. We note that the crystals grow directly in the tetragonal phase at these temperatures, resulting in high-quality single crystals free from structural transformations or decomposition reactions.  Wavelength dispersive X-ray spectroscopy reveals an impurity level below 500 ppm.  In particular,  there is no evidence of  Cl, Si, K or Al impurities.  X-ray diffraction confirms the tetragonal structure with lattice constants $a$ = 3.7707(12) \AA~ and $c$ = 5.521(3) \AA. Structural refinement shows the stoichiometric composition of Fe and Se within the error (Fe:Se=0.995(4):1). 
The structural $z$ parameter of Se is $z_{\rm Se} = 0.26668(9)$. No indications for interstitial atoms were found.

The extremely small level of  impurities and defects (less than one impurity per 2000 Fe atoms) are confirmed by scanning tunneling microscope (STM) topography (Fig.\;S1).

Figures\;S2{\it B} shows the magnetic susceptibility $\chi$ of the sample used for the penetration depth measurements (Fig.\;1{\it B}), which is measured by superconducting quantum interference device magnetometer under zero-field-cooling condition with a field applied parallel to the $c$ axis  ($H=1$\;Oe).   The sample size is approximately $300\times300\times10$ $\mu$m$^3$.  The  susceptibility shows sharp transition with the width (10-90\%) of 0.3\;K and  $T_{\rm c}$  defined as the mid-point of the transition is 9.25\;K.    Figure\;S2{\it A} shows the resistive transition of another sample of the same batch (inset of Fig.\;1{\it A}).    The temperature at which the resistivity goes to zero is 9.40\;K, which is very close to $T_{\rm c}$ determined by $\chi$.

The high quality of our samples allow observing quantum oscillations above $H_{\rm irr}$ \cite{Tera14}. 
 The results, together with large $RRR$ value (Fig.\;1{\it A}),  large magetoresistance (inset of Fig.\;1{\it A})  and extremely small level of  impurities and defects (Fig.\;S1),  demonstrate  that the crystals used in the present study are very clean.  

\begin{figure}[h]
\begin{center}
\includegraphics[width=0.7\linewidth]{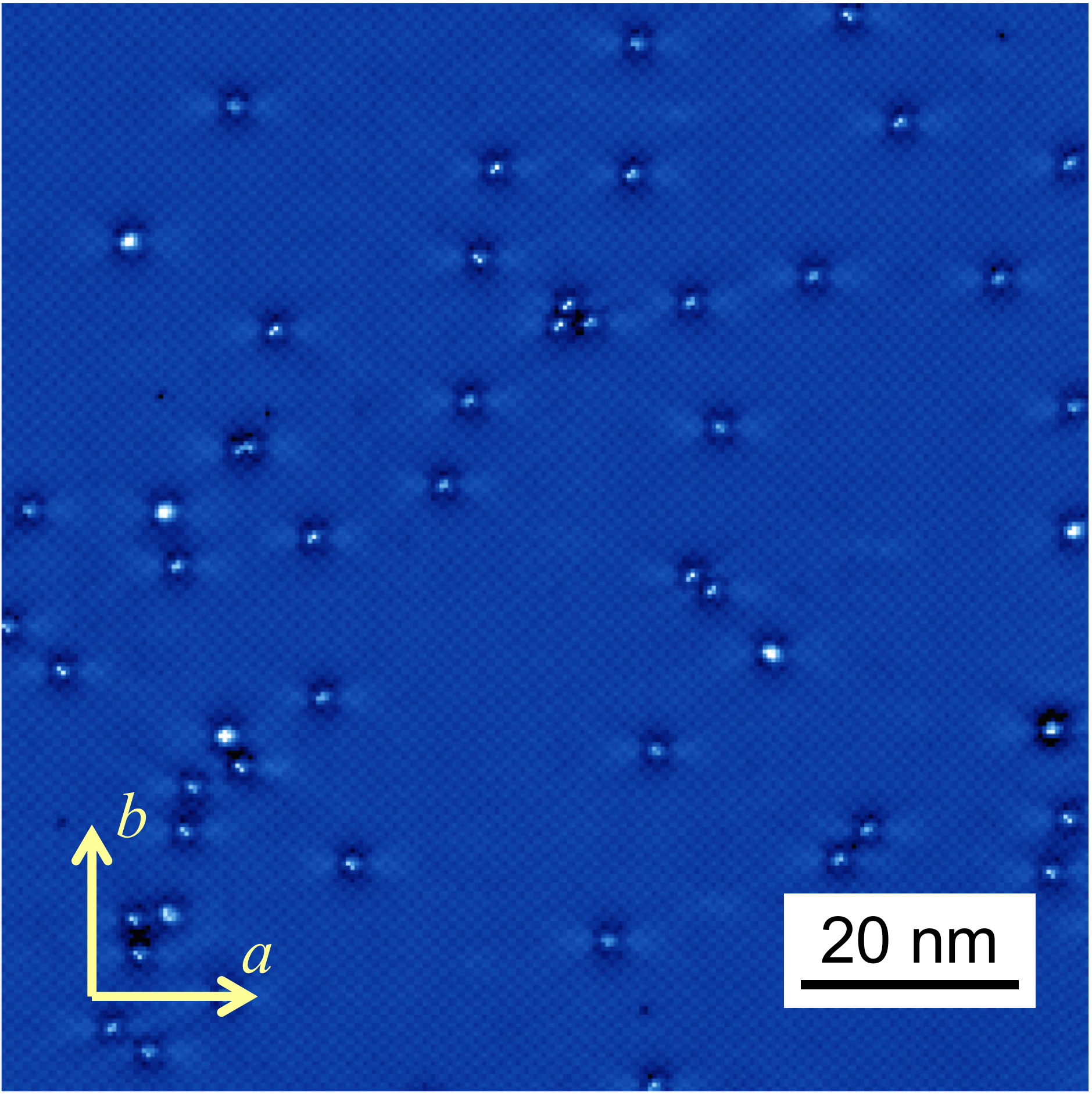}
\end{center}
\vspace{6pt}
\caption{STM topograph of FeSe at 1.5\;K. White bright spots are impurities or defects. Feedback conditions are sample bias voltage $V_{\rm s}=$+95\;mV and tunneling current $I_{\rm t}$=10\;pA.} 
\end{figure}
\begin{figure}[h]
\begin{center}
\centerline{\includegraphics[width=.59\linewidth]{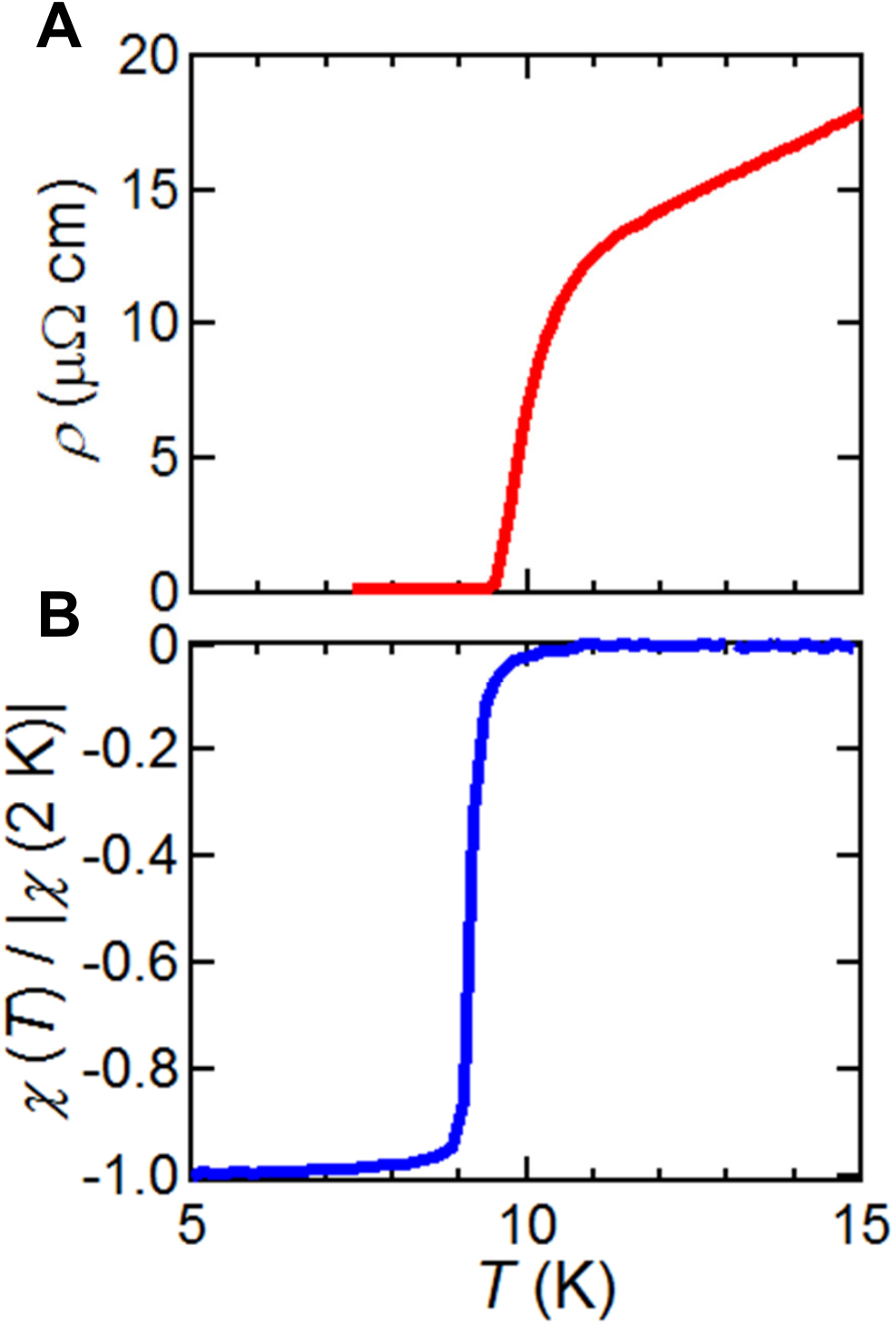}}
\end{center}
\caption{Superconducting transition of FeSe single crystals. ({\it A}) Resistive transition in zero field.  ({\it B}) Magnetic susceptibility measured under zero-field-cooling condition.} 
\end{figure}

\section{SI2--Transport measurements}

Thermal conductivity and magnetoresistance were measured on the same crystal  ($880\times340\times10$ $\mu$m$^3$) using the same contacts.  We attached the contacts after cleaving the surface.  The thermal conductivity was measured by the standard steady-state method in a dilution refrigerator.   Hall and Seebeck measurements were performed on another crystal  ($850\times1000\times5$ $\mu$m$^3$).  

Above 10\;K, the magnetoresistance $\Delta \rho/\rho\equiv (\rho(H)-\rho)/\rho$ exhibits a $H^n$ ($n \sim 2)$ dependence without saturation (Fig.\;S3{\it A}), which demonstrates the nearly perfect compensation, i.e., an equal density of electrons and holes, $n_{\rm h}=n_{\rm e}$.  In a compensated metal, the magnetoresistance is given by $ \Delta \rho/\rho=(\omega_{\rm c}^{\rm e}\tau_{\rm e})(\omega_{\rm c}^{\rm h}\tau_{\rm h})$, where $\omega_{\rm c}=e \mu H/m$ is the cyclotron frequency for carriers with mass $m$ and scattering time $\tau$ \cite{Pip89}.  The suffixes e and h denote ``electron" and ``hole", respectively.   At $T=10$\;K we estimate  $(\omega_{\rm c}^{\rm e} \tau_{\rm e})(\omega_{\rm c}^{\rm h} \tau_{\rm h}) \approx 5$ at 10\;T, indicating the high mobility of charge carriers.  

Figure\;S3{\it B} shows the temperature dependence of the Hall coefficient $R_{\rm h}$  in the zero-field limit.  Above 100\;K, $R_{\rm h}$ is close to zero.  Below $\sim$ 60\;K, $R_{\rm h}$ is negative and strongly temperature dependent.  In a compensated metal, the  Hall coefficient is given by  $R_{\rm h}=\frac{1}{n_{\rm e}e}\frac{\omega_{\rm c}^{\rm e}\tau_{\rm e}-\omega_{\rm c}^{\rm h}\tau_{\rm h}}{\omega_{\rm c}^{\rm e}\tau_{\rm e}+\omega_{\rm c}^{\rm h}\tau_{\rm h}}$ \cite{Pip89}.  The strong $T$ dependence of $R_{\rm h}$ indicates that the electron and hole mobilities are of the same order \cite{Kas07}, which is consistent with the QPI results. 

Figure\;S3{\it C} shows the temperature dependence of the Seebeck coefficient divided by $T$, $S/T$.  Below 40\;K, $S$ is negative.   The Seebeck coefficient  in the single-band case is expected to be $T$-linear in the zero-temperature limit and linked to $\varepsilon_{\rm F}$ by $S/T=\pm \frac{\pi^2}{2}\frac{k_{\rm B}^2}{e}\frac{1}{\varepsilon_{\rm F}}$.  In a multiband system with both electrons and holes contributing with opposite signs to the overall Seebeck response, the single-band formula sets an upper limit to the Femi energy of the dominant band \cite{Pour11}.  From $S/T\sim 3.5$\;$\mu$V/K$^2$ at low temperatures above $T_{\rm c}$,  we estimate  the upper limit of $\varepsilon_{\rm F}^{\rm e}$ to be $\sim$10\;meV. 

\begin{figure*}[t]
\begin{center}
\includegraphics[width=0.9\linewidth]{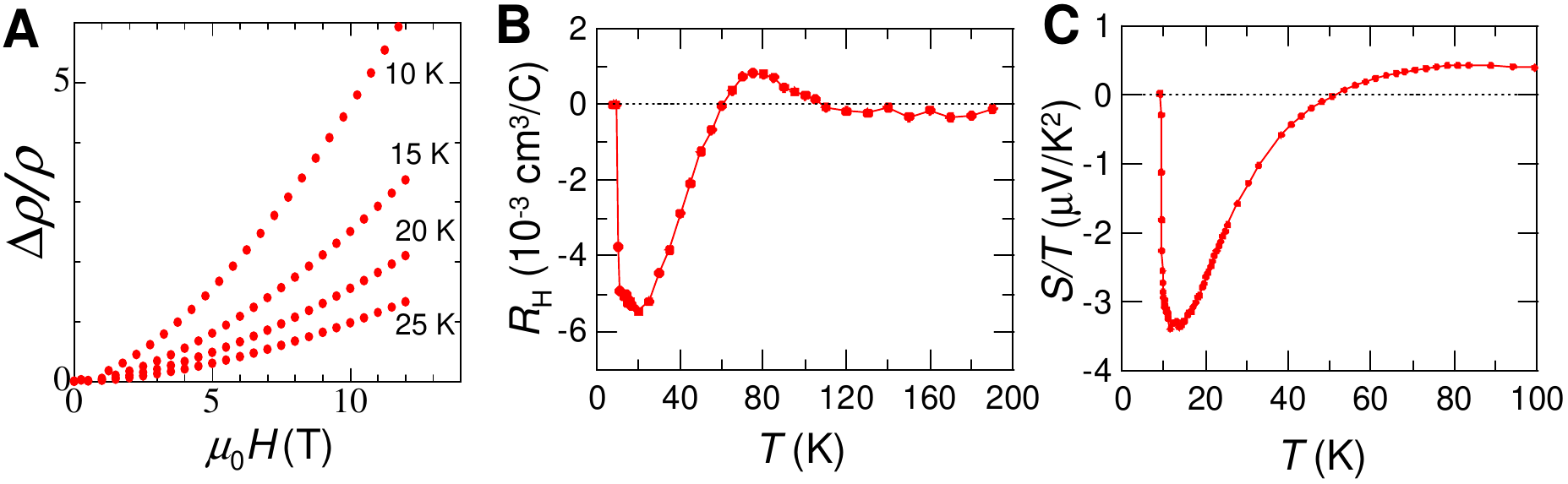}
\end{center}
\caption{Transport properties of FeSe single crystals. ({\it A}) Magnetoresistance above 10\;K.  ({\it B}) Temperature dependence of the Hall coefficient $R_{\rm H}$.  ({\it C}) Temperature dependence of the Seebeck coefficient divided by $T$, $S/T$.}
\end{figure*}


\section{SI3--London penetration depth}

 To determine the absolute value of the in-plane London penetration depth $\lambda_{\rm L}(0)$ in a small single crystal reliably, we combined the high-precision tunnel diode oscillator (TDO) (resonant frequency of $f=$\;13\;MHz) and the microwave cavity perturbation  ($f=$\;28\;GHz) techniques \cite{Hash12}.

For the TDO technique, we can determine the change of the London penetration depth $\delta \lambda_{\rm L}\equiv \lambda_{\rm L}(T)-\lambda_{\rm L}(0)$ by the change of the resonant frequency $\delta f\equiv f(T)-f(0)$, $\delta f=G \delta \lambda_{\rm L}$. The calibration factor $G$ is determined from the geometry of the sample.  We measured $\delta f$ down to 100\;mK. 

For the microwave cavity perturbation technique, we used a superconducting cavity resonator with high $Q$-factor ($Q>10^6$).   We measured the microwave surface impedance $Z_{\rm s}=R_{\rm s}+iX_{\rm s}$ in the Meissner state down to 4.2\,K, where $R_{\rm s}$ and $X_{\rm s}$ are the surface resistance and  reactance, respectively (Fig.\;S4{\it A}).  In the present frequency range, the complex conductivity $\sigma = \sigma_1- i\sigma_2$  in the skin-depth regime is given by $Z_{\rm s}$ through the relation:
\begin{eqnarray}
Z_{\rm s} = R_{\rm s} + iX_{\rm s} = \left(\frac{i\mu_0\omega}{\sigma_1 - i\sigma_2}\right)^{1/2}
\label{ComplexConductivity_eq}.
\end{eqnarray}
In the Hagen-Rubens limit, $\omega\tau \ll 1$, where $\omega$  is the microwave frequency and  $\tau$ is the scattering rate, $\sigma_2$ is related to $\lambda_{\rm L}$ by
$\sigma_2 = \frac{1}{\mu_0\omega\lambda_{\rm L}^2}$. 
In Fig.\;S4{\it B}, the blue circles show $\sigma_2(T)/\sigma_2(4.2$\;K), which represent the normalized superfluid density $\rho_{\rm s}$.   The solid lines represent $\rho_{\rm s}$ obtained from $\delta \lambda_{\rm L}(T)$ assuming several different $\lambda_{\rm L}(0)$ values.   The best fit is obtained for $\lambda_{\rm L}(0)\approx 400$\;nm.

\begin{figure*}[h]
\begin{center}
\includegraphics[width=0.7\linewidth]{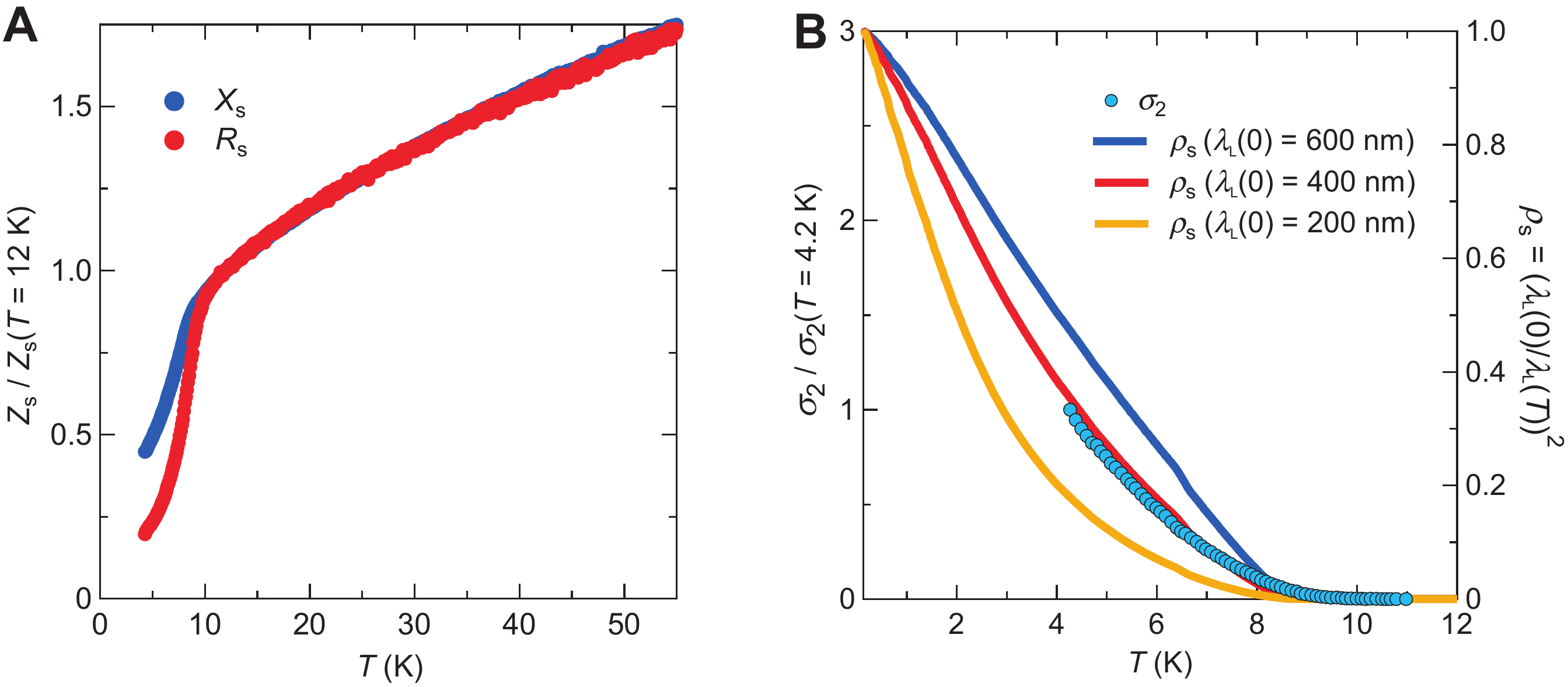}
\end{center}
\caption{Determination of absolute value of the penetration depth. ({\it A}) Temperature dependence of the microwave surface resistance $R_{\rm s}$ and reactance $X_{\rm s}$. ({\it B}) Temperature dependence of $\sigma_2$ in microwave surface impedance measurements (blue circles, right axis) and normalized superfluid density $\rho_{\rm s}$ calculated from $\delta\lambda_{\rm L}$ in TDO measurements by assuming different values of $\lambda_{\rm L}(0)$ (solid lines).  } 
\end{figure*}

\vspace{-1cm}
\section{SI4--Quasiparticle interference}
Figures\;S5{\it A-L} show the energy dependent QPI patterns at 12~T. 
Figures\;S5{\it A-C} display the normalized conductance images of occupied states. In order to avoid the so-called set-point effect associated with the spatial variation of the integrated density of states~\cite{Chen_book}, raw conductance data $dI/dV$ is normalized by $I/V$, where $I$ and $V$ are tunneling current and bias voltage, respectively~\cite{Feenstra}. The nearest-neighbor Fe-Fe distance is larger along the $b$ axis than along the $a$ axis.  Figures\;S5{\it D-F} display Fourier-transformed images of Figures\;S5{\it A-C}. Figures\;S5{\it G-I} display the normalized conductance images of unoccupied  states.  Figures\;S5{\it J-L} display Fourier-transformed images of Figures\;S5{\it G-I}.

Figure\;S6{\it A} shows STM topographic image of the dumbbell-shaped impurity.  Figure\;S6{\it B} shows STS spectra at the impurity (blue) and at the position far from the impurity (red).  A sharp peak at $+10$\;meV in the STS spectrum (blue) arises from the impurity bound state, which gives rise to the $q$-independent dispersion shown in Figs.\;2{\it B-E}.

\begin{figure*}[t]
\begin{center}
\includegraphics[width=0.9\linewidth]{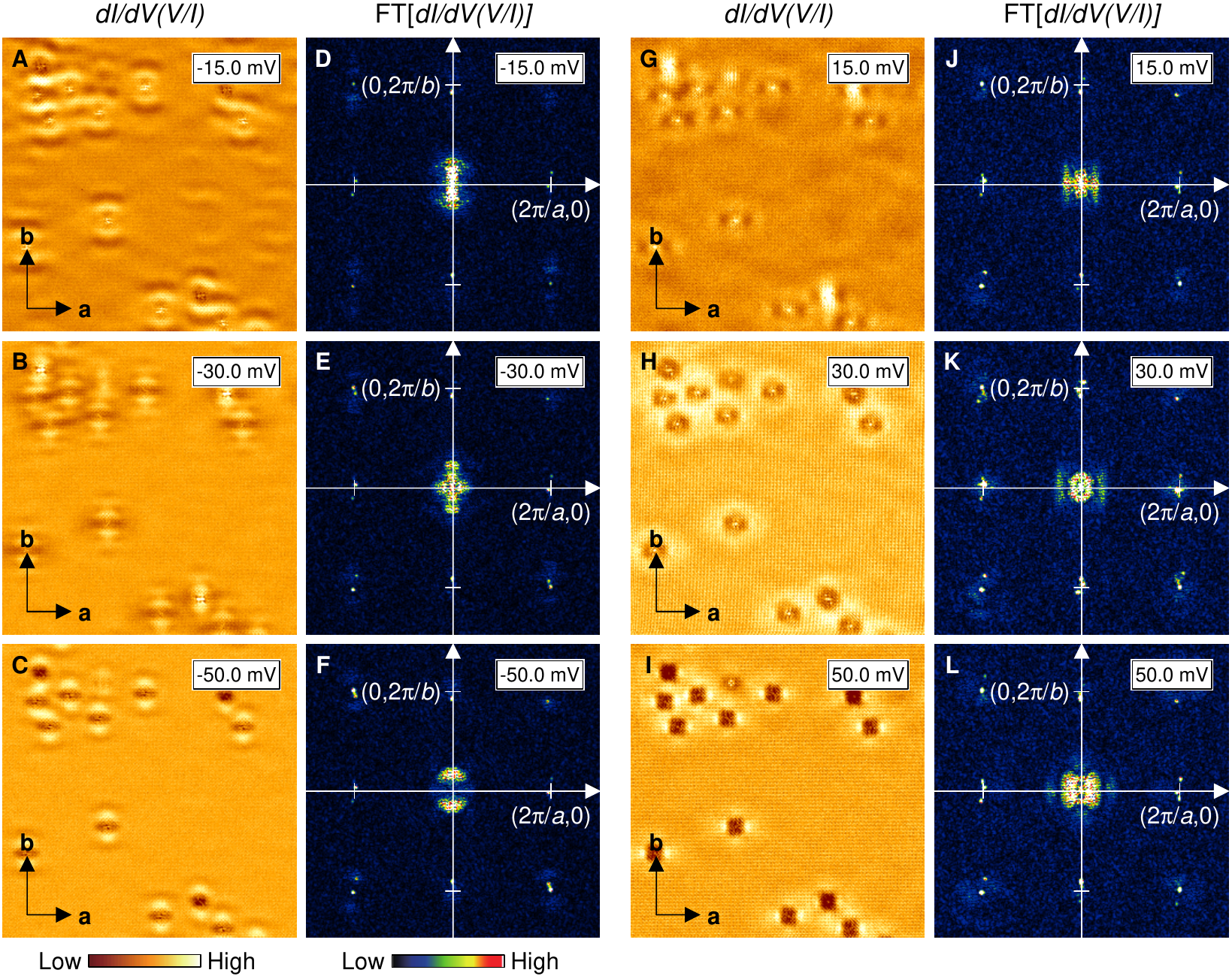}
\end{center}
\caption{Energy dependent QPI patterns at $T=1.5$\,K and $\mu_0H=12$\;T.
Scan area is 45\;nm$\times$45\;nm.  Feedback conditions are $V_{\rm s}=+50$\;mV and $I_{\rm t}=100$\;pA.  Bias modulation amplitude for spectroscopy was set to 1\;mV$_{\rm rms}$. (A-C) Normalized conductance, $dI/dV(V/I)$, images of the occupied states at $V=-15$ mV (A), -30 mV (B), -50 mV (C). (D-F) Fourier transform of the images shown in (A)-(C). (G-I) Normalized conductance images of the empty states at $V=+15$ mV (G), +30 mV (H), +50 mV (I). (J-L) Fourier transform of the images shown in (G)-(I). } 
\end{figure*}


\begin{figure*}[t]
\begin{center}
\includegraphics[width=0.55\linewidth]{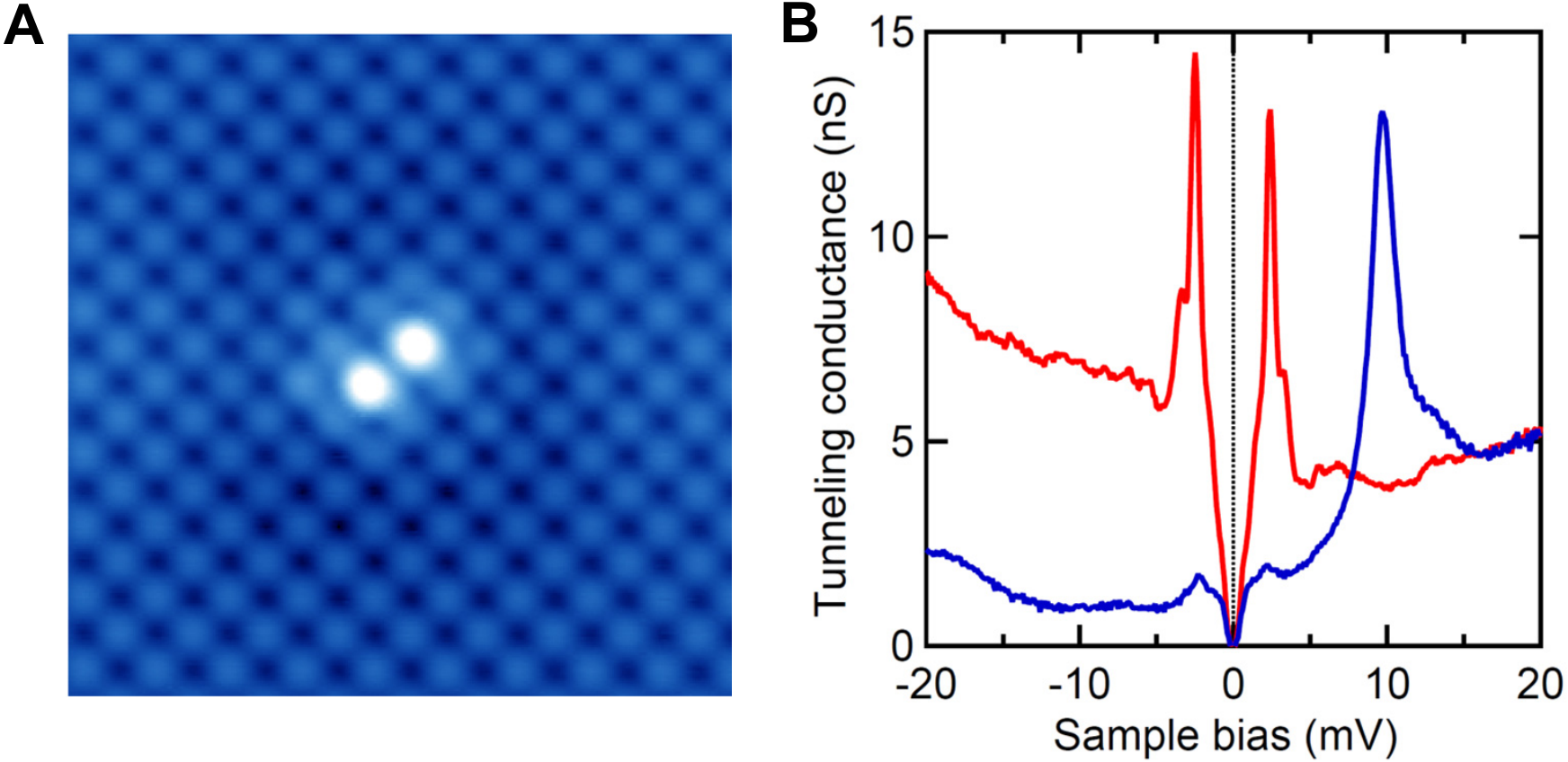}
\end{center}
\caption{Impurity states revealed by STM. ({\it A}) 5 $\times$ 5 nm$^2$ STM topographic image of the dumbbell-shaped impurity taken at $T =0.4$\;K. Feedback conditions are sample bias voltage $V_{\rm s}=+95$\;mV and tunneling current $I_{\rm t}=100$\;pA. ({\it B}) Tunneling conductance spectra at $T =0.4$\;K. Spectra were taken with bias modulation amplitude $50~\mu{\rm V}_{rms}$ after freezing the tip at $V_{\rm s}$ =+20~mV and $I_t$ = 100~pA.
} 
\end{figure*}

\section{SI5--Magnetic torque}
The magnetic torque was measured using a piezo-resistive micro-cantilever technique down to 30\;mK and up to 17.8\;T.   A small single crystal of approximately $100\times100\times15$ $\mu$m$^3$ was mounted on the lever with a tiny amount of Apiezon grease. The field is slightly tilted away from the $c$ axis.  Figure\;S7 shows the field dependence of the torque signal.  The irreversibility field $H_{\rm irr}$ is defined by the point where the hysteresis loop has closed to a level of 0.3\% (arrows pointing down).  We note that $H_{\rm irr}$ determined by the magnetic torque coincides well with the $H_{\rm irr}$ defined by the zero resistivity as shown in Fig.\;4 in the main text.   A broad peak effect associated with the order-disorder transition of the flux-line lattice is observed  after subtraction of a smooth background (arrows pointing up indicate the maximum).  The peak field is seen to be strongly temperature dependent,  in contrast to the $H^{\ast}$ line.  
\begin{figure}[b]
\begin{center}
\includegraphics[width=0.65\linewidth]{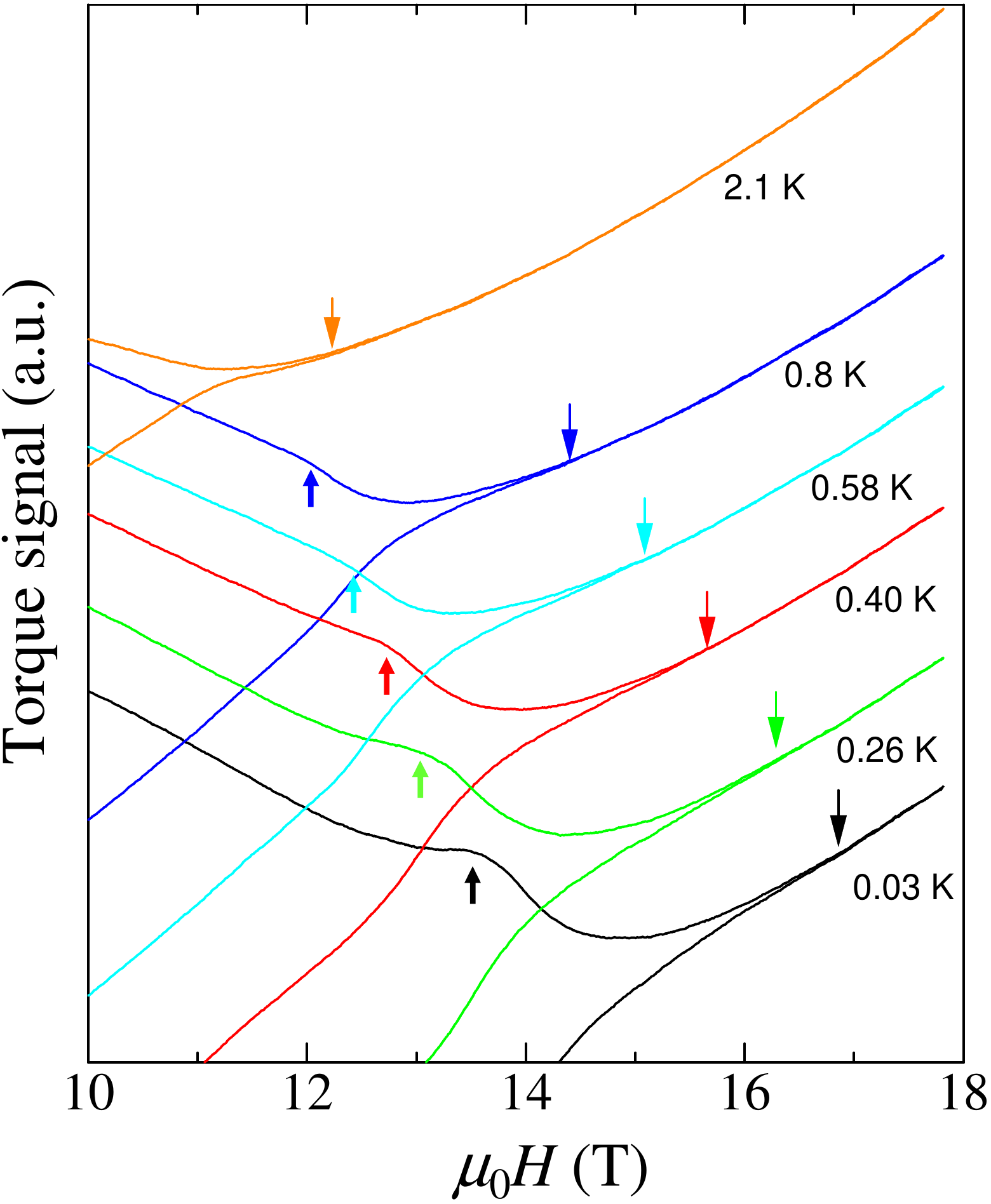}
\end{center}
\caption{Field dependence of the magnetic torque at low temperatures. Each curve is vertically shifted for clarity. Downward (upward) arrows mark the positions of the irreversibility (peak) field.} 
\end{figure}

\end{article}

\end{document}